\documentclass[pra,twocolumn,superscriptaddress] {revtex4-2}
\usepackage{graphicx}
\usepackage{amssymb}
\usepackage{amsfonts}
\usepackage{amsmath}
\usepackage{lmodern}
\usepackage{mathrsfs}
\usepackage{mathtools, slashed}
\usepackage{mathdots}
\usepackage{collref}
\usepackage[colorlinks=true, citecolor=blue, urlcolor=blue]{hyperref}
\usepackage{tikz}
\usepackage{bm}
\usepackage[markup=underlined]{changes}
\makeatletter
\AddToHook{cmd/added/before}{\def\Changes@AuthorColor{purple}}
\AddToHook{cmd/deleted/before}{\def\Changes@AuthorColor{green}}
\AddToHook{cmd/replaced/before}{\def\Changes@AuthorColor{red}}
\makeatother
\usepackage{todonotes}
\setcommentmarkup{\todo[color={pink}, size=\scriptsize]{#3: #1}}

\usetikzlibrary{shapes.geometric}

\renewcommand{\section}[1]{{\par\it #1.---}\ignorespaces}
\begin{document}
\title{Underlying orbital U(1) symmetry in quantum limit cycles}
\author{Chong Chen}
\email{chongchenn@gmail.com}
\affiliation{Department of Physics, The Chinese University of Hong Kong, Shatin, New Territories, Hong Kong, China}
\date{\today}

\begin{abstract}
Quantum limit cycles exhibit harmonic Liouvillian spectra, yet their underlying orbital $U(1)$ structure is not generally explicit in the absence of physical rotational symmetry. Here, we construct a unitary representation of orbital phase translations through a similarity transformation that maps the harmonic nondecaying Liouvillian sector onto a Hermitian angular-momentum Hamiltonian. The resulting phase states possess homogeneous Fubini--Study geometry independently of the physical orbit's shape. For ideal quantum limit cycles, the corresponding  density matrix also exhibit homogeneous Bures geometry. At finite size, we show how phase diffusion damps slow Liouvillian harmonics while preserving the effective orbital $U(1)$ symmetry. The resulting loss of phase-state distinguishability and contraction of the effective geometry are quantitatively predicted by the microscopic slow Liouvillian spectrum. Our results establish a connection between orbital phase symmetry, quantum geometry, and finite-size phase diffusion in open quantum systems.
\end{abstract}

\maketitle

\section{Introduction}
A stable limit cycle is an isolated periodic orbit that attracts nearby trajectories. In driven-dissipative systems, such periodic attractors arise from the interplay between pumping and dissipation~\cite{Strogatz2024}. This concept extends naturally to open quantum systems with pumping and dissipation, where it manifests as a periodic asymptotic state with sustained oscillations in observables~\cite{Loerch2014, NavarreteBenlloch2017, BenArosh2021, Nadolny2026}. Unlike equilibrium steady states, such dynamical attractors encode a well-defined phase degree of freedom and exhibit robustness against perturbations. These features place quantum limit cycles in close connection with quantum synchronization~\cite{Lee2013,Walter2014,Kehrer2025,Nadolny2026} and spontaneous time-translation symmetry breaking \cite{Wilczek2012, Bruno2013, Watanabe2015, Kessler2019, Zaletel2023}, providing a natural platform for realizing continuous time crystals \cite{Iemini2018, Kessler2021, Kongkhambut2022, Wu2024}.

The Liouvillian formalism~\cite{AlbertJiang2014, AlbertBradlyn2016} provides a microscopic description of quantum limit cycles, revealing the algebraic structure of nonequilibrium steady states~\cite{Burgarth2013}, the conditions for persistent oscillations~\cite{Buca2019,Buca2022}, and their connection to emergent coherence and dynamical symmetries~\cite{Dutta2025,Daviet2024,Daviet2025}. In particular, Liouvillian eigenmodes with purely imaginary eigenvalues give rise to persistent oscillations~\cite{Buca2019,Buca2022}. This connects quantum limit cycles to the broader theory of decoherence-free dynamics~\cite{Lidar1998}. Ref.~\cite{Dutta2025} identified an approximately parabolic branch of long-lived Liouvillian modes that approaches an equally spaced imaginary spectrum in the classical limit, linking finite-size damping to phase diffusion. Quantum asymptotic phases have also been constructed from adjoint Liouvillian eigenoperators~\cite{KatoNakao2022,KatoNakao2023}. Together, these studies provide a spectral framework for understanding persistent oscillations and their finite-size relaxation.

Previous studies have established the spectral signatures of quantum limit cycles and characterized persistent oscillations through Liouvillian dynamical symmetries~\cite{Buca2019,Buca2022,KatoNakao2022,KatoNakao2023,Dutta2025}. However, these spectral results do not by themselves specify the underlying orbital $U(1)$ symmetry and the quantum geometry for orbital phase states. For noncircular limit cycles, this question is particularly relevant because translations along the orbit cannot generally be identified with rotations in physical phase space. A central challenge is therefore to construct an effective phase-state representation directly from the Liouvillian spectrum and to characterize how its geometry responds to finite-size relaxation.

In this Letter,  we recover the underlying orbital $U(1)$ structure from the harmonic Liouvillian sector without assuming rotational symmetry in physical phase space. Through a similarity transformation, we map this sector onto a Hermitian angular-momentum Hamiltonian, explicitly realizing the dynamical symmetry as orbital phase translations. The resulting phase-related states possess homogeneous Fubini--Study geometry, even for noncircular limit cycles. In the ideal limit, the corresponding density matrix also exhibit homogeneous Bures geometry, while the stationary quantum state displays a nearly uniform intrinsic orbital-phase distribution even at finite size. This orbital symmetry further provides a natural framework for understanding finite-size phase diffusion, in close analogy with fluctuations in systems exhibiting spontaneous continuous-symmetry breaking. In the semiclassical weak-noise regime, restoring dynamics suppresses transverse fluctuations, whereas fluctuations along the neutrally stable phase direction accumulate, leading to phase diffusion and approximately quadratic damping of the slow Liouvillian harmonics. Such relaxation preserves the effective orbital $U(1)$ covariance while progressively reducing phase-state distinguishability. The resulting contraction of the effective geometry is quantitatively predicted by the microscopic slow Liouvillian spectrum. Our results thus connect the spectral characterization of quantum limit cycles with their underlying orbital symmetry, quantum geometry, and finite-size relaxation.

\begin{figure}[thpb]
    \centering
    \includegraphics[width=1.0\linewidth]{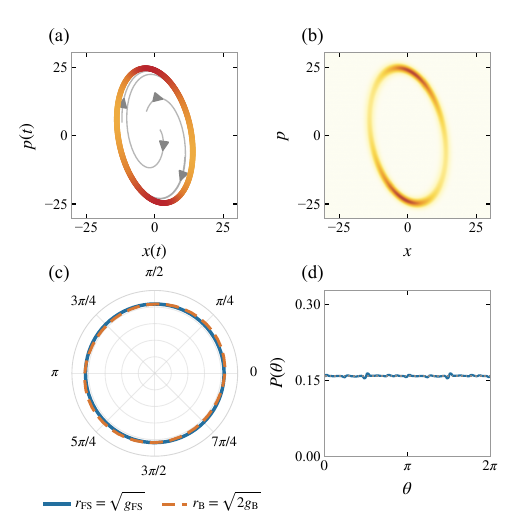}
    \caption{ Classical, quantum, and geometric representations of a noncircular cycle. (a)  classical limit cycle obtained from the mean-field equations, with color indicating the normalized orbital weight. (b) Husimi-Q distribution of the quantum stationary state, exhibiting a noncircular phase-space structure corresponding to the classical orbit. (c) Effective Fubini--Study radius $r_{\mathrm{FS}}=\sqrt{g_{\mathrm{FS}}}$ (solid) and rescaled Bures radius $r_{\mathrm B}=\sqrt{2g_{\mathrm B}}$ (dashed) at $\omega_0t=5.0$, showing homogeneous effective geometry and weak finite-size modulation, respectively. (d) Intrinsic orbital-phase distribution $P(\theta)=\int d^2\alpha\,Q(\alpha)\delta(\theta-\theta(\alpha))$, exhibiting nearly uniform phase statistics, $P(\theta)\simeq1/(2\pi)$ (illustrated by gray dashed line), despite the noncircular physical orbit.  Parameters are $\omega_0=1.0$,  $\epsilon/\omega_0=0.6$, $\gamma_\uparrow/\omega_0=1.0$,$\gamma_\downarrow/\omega_0=0.2$, $\kappa/\omega_0=0.002$, with the initial coherent state evolved for $\omega_0 \tau=1.5$.}
    \label{fig:limitCycle}
\end{figure}

\section{Classical limit cycle}
A quantum limit cycle often admits a classical counterpart described by the corresponding mean-field dynamics. As a representative example, we consider a modified quantum van der Pol oscillator introduced in Ref.~\cite{Walter2014, Sonar2018, BenArosh2021}, governed by the master equation
\begin{equation}\label{eq:masterEquation}
    \frac{d\rho(t)}{dt}
    =-i[\hat H_0,\rho(t)]
    +\gamma_{\uparrow}\mathcal D_{\hat a^\dagger}[\rho]
    +\gamma_{\downarrow}\mathcal D_{\hat a}[\rho]
    +\kappa \mathcal D_{\hat{a}^2 }[\rho],
\end{equation}
where $\hat H_0=\omega_0\hat a^\dagger\hat a+\frac{\epsilon}{2}(\hat a^2+\hat a^{\dagger2})$, and $\mathcal D_{\hat o}[\rho]=\hat o\rho\hat o^\dagger-\{\hat o^\dagger\hat o,\rho\}/2$ denotes the Lindblad dissipator associated with jump operator $\hat o$. Here, $\hat a$ and $\hat a^\dagger$ are bosonic annihilation and creation operators, while $\hat x=(\hat a+\hat a^\dagger)/\sqrt{2}$ and $\hat p=i(\hat a^\dagger-\hat a)/\sqrt{2}$ denote the quadratures.  The dissipative terms describe linear pumping, linear damping, and nonlinear damping, respectively, while the quadratic Hamiltonian term $\frac{\epsilon}{2}(\hat a^2+\hat a^{\dagger2})$ introduces anisotropy into the dynamics.

Defining $x(t)=\mathrm{Tr}[\hat x\rho(t)]$ and $p(t)=\mathrm{Tr}[\hat p\rho(t)]$, and applying the mean-field factorization $\langle\prod_i\hat o_i\rangle\simeq\prod_i\langle\hat o_i\rangle$, we obtain
\begin{align}
    \dot x(t)&=\left[\frac{\gamma_\uparrow-\gamma_\downarrow}{2}
    -\frac{\kappa}{2}(x^2+p^2)\right]x(t)+(\omega_0-\epsilon) p(t),\nonumber\\
    \dot p(t)&=\left[\frac{\gamma_\uparrow-\gamma_\downarrow}{2}
    -\frac{\kappa}{2}(x^2+p^2)\right]p(t)-(\omega_0+\epsilon) x(t).
    \label{eq:MeanField}
\end{align}
 For $\omega_0>|\epsilon|$, the normal solution $x=p=0$ is stable when $\gamma_\uparrow<\gamma_\downarrow$. Increasing the pumping rate through $\gamma_\uparrow=\gamma_\downarrow$ induces a Hopf bifurcation, giving rise to a stable limit cycle~\cite{Strogatz2024}.  The Hamiltonian anisotropy deforms the orbit and makes its angular velocity nonuniform. The orbital frequency can be determined as $\omega=\sqrt{\omega_0^2-\epsilon^2}$ (See Appendix~\ref{sec:Reparameterization-App}).

Figure~\ref{fig:limitCycle}(a) illustrates the resulting classical limit cycle. The color encodes the normalized orbital weight $w(x,p)=[T \sqrt{\dot x^2(t)+\dot p^2(t)}]^{-1}$, which satisfies $\oint_{\mathcal C}w(x,p)dl=1$, where $dl=\sqrt{dx^2+dp^2}$ and $T=2\pi/\omega$ is the oscillation period. This weight describes the fraction of time spent along different segments of the orbit. 

\section{Quantum limit cycle}
An ideal quantum limit cycle exhibits persistent periodic evolution of the density matrix after transient relaxation. To characterize this behavior, consider the master equation $\dot\rho=\mathcal L\rho$, with the density matrix vectorized as $|\rho\rangle\rangle$. Assuming that the Liouvillian is diagonalizable on the relevant operator space, its spectral decomposition gives
\begin{equation}
    \mathcal{L}=\sum_{j} z_j |\rho_{j}\rangle \rangle \langle \langle \tilde{\rho}_{j}|,~~ |\rho(t)\rangle\rangle =\sum_j c_j e^{z_j t}  |\rho_{j}\rangle \rangle,
\end{equation}
where $|\rho_j\rangle\rangle$ and $\langle \langle \tilde{\rho}_j|$ are biorthogonal right and left eigenmodes with eigenvalues $z_j\equiv -\gamma_j -i \epsilon_j$, and $c_j=\langle \langle \tilde{\rho}_{j}|\rho(0)\rangle\rangle$. For a Lindblad generator, $\gamma_j\geq0$, and at least one zero eigenvalue corresponds to a stationary state~\cite{Kessler2012, Minganti2018}.

While decaying modes determine transient relaxation, the long-time dynamics is governed by the nondecaying Liouvillian sector~\cite{Buca2019,Buca2022}. Persistent oscillations require populated nonstationary modes with $\gamma_l=0$ and $\epsilon_l\neq0$. After the decaying modes have relaxed, the density matrix approaches
\begin{equation}\label{eq:limitCycle}
|\rho(t)\rangle\rangle
\simeq\sum_l c_l e^{-i\epsilon_lt}|\rho_l\rangle\rangle,
\qquad \gamma_{g}t\gg1,
\end{equation}
where the sum includes the stationary component with $\epsilon_0=0$, and $\gamma_{g}$ characterizes the relaxation of decaying modes. For a periodic trajectory with minimal period $T$, the populated frequencies are integer multiples of a fundamental frequency, $\epsilon_l=l\omega$, where $l\in\mathbb Z$ and $\omega=2\pi/T$~\cite{Dutta2025}. The resulting asymptotic dynamics is periodic rather than stationary.

For a phase-selected persistent cycle, a generic time translation changes the state, whereas translation by an integer number of periods leaves it invariant:
\begin{equation}
\rho(t+\tau)\neq\rho(t),
\quad
\rho(t+nT)=\rho(t),
\end{equation}
for generic $\tau$ and integer $n$.
This is the dynamical structure associated with spontaneous breaking of continuous time-translation symmetry to discrete translations by $T$~\cite{Wilczek2012,Iemini2018}. The distinct orbital phase origins are therefore parameterized by the quotient of the continuous time-translation group by the discrete subgroup of period translations, given by $\mathbb R/(T\mathbb Z)\simeq U(1)$~\cite{Daviet2025}. Importantly, this orbital $U(1)$ structure does not require the Liouvillian $\mathcal L$ to possess physical rotational symmetry. Its explicit representation in Liouville space is constructed below.

At finite size, the oscillatory modes generally acquire small decay rates, giving rise to long-lived oscillations followed by phase relaxation~\cite{Macieszczak2016}.   Nevertheless, the stationary-state Husimi--$Q$ distribution can retain a clear signature of the underlying orbit~\cite{Carmichael1999}. In Fig.~\ref{fig:limitCycle}(b), we illustrate the Husimi--$Q$ distribution of the quantum stationary state $\rho_s$, satisfying $\mathcal{L} \rho_s=0$. It exhibits a noncircular ring-like structure closely resembling the classical limit cycle.

\section{Underlying $U(1)$ symmetry}
The harmonic Liouvillian spectrum of an ideal quantum limit cycle encodes a dynamical $U(1)$ structure associated with periodic evolution. However, this structure need not correspond to rotations in physical phase space, and its unitary representation is obscured by the generally nonorthogonal Liouvillian eigenmodes. To make this structure explicit, we consider the nondecaying subspace~\cite{Buca2019}
\begin{equation}
\mathcal S_{\mathrm{DFS}}
=\operatorname{span}\{|\rho_l\rangle\rangle:\gamma_l=0\},
\quad
\mathcal L_{\mathrm{DFS}}
=\sum_l(-il\omega)\mathcal P_l,
\label{eq:LDFS}
\end{equation}
where $\mathcal P_l=|\rho_l\rangle\rangle\langle\langle\tilde\rho_l|$ are the biorthogonal spectral projectors. Although $\mathcal L_{\mathrm{DFS}}$ is generally non-Hermitian in the Hilbert--Schmidt inner product, it can be mapped to an anti-Hermitian generator through a similarity transformation. Defining the positive operator
\begin{equation}
\eta=\sum_l
\frac{|\rho_l\rangle\rangle\langle\langle\rho_l|}
{\langle\langle\rho_l|\rho_l\rangle\rangle}
\end{equation}
on the nondecaying sector gives
$\mathcal L_{\mathrm{DFS}}\eta+\eta\mathcal L_{\mathrm{DFS}}^\dagger=0$.
This definition is invariant under arbitrary nonzero rescalings of the Liouvillian eigenmodes. Consequently,
\begin{equation}
H\equiv i\eta^{-1/2}\mathcal L_{\mathrm{DFS}}\eta^{1/2}
=\omega\sum_l l|l\rangle\langle l|
\equiv\omega\hat L,
\label{eq:Heff}
\end{equation}
where $|l\rangle=(Z_l\eta)^{-1/2}|\rho_l\rangle\rangle$, with $Z_l=\langle\langle\rho_l|\rho_l\rangle\rangle$, form an orthonormal basis. Here, $\eta^{-1}$ defines the inner product on the nondecaying sector, satisfying $\langle\langle\rho_l|\eta^{-1}|\rho_m\rangle\rangle=Z_l \delta_{l,m}$.

The Hermitian angular-momentum operator $\hat L$ generates a unitary $U(1)$ representation,
\begin{equation}
U(\phi)=e^{-i\phi\hat L},
\quad
e^{-iHt}=U(\omega t),
\label{eq:U1}
\end{equation}
with $U(\phi+2\pi)=U(\phi)$. Thus, the effective time evolution is itself a $U(1)$ phase translation, rather than merely being invariant under one. In the original Liouville space, the corresponding transformations are $\mathcal U(\phi)=\eta^{1/2}U(\phi)\eta^{-1/2}=\exp(\phi\mathcal L_{\mathrm{DFS}}/\omega)$. The similarity transformation therefore recovers a unitary representation of the dynamical $U(1)$ structure.

To establish its orbital interpretation, we map the density matrix $|\rho(\theta)\rangle\rangle$ along the limit cycle onto normalized states in the effective Hermitian representation,
\begin{equation}
|\psi(\theta)\rangle
=\frac{\eta^{-1/2}|\rho(\theta)\rangle\rangle}
{\sqrt{\langle\langle\rho(\theta)|\eta^{-1}|\rho(\theta)\rangle\rangle}}
=\sum_l\sqrt{p_l}\,e^{il\theta}|l\rangle,
\label{eq:PhaseStates}
\end{equation}
where $p_l$ denotes the normalized weight of the $l$-th harmonic, with the eigenmode phases fixed relative to a reference state. These states form a coherent-state-like family satisfying
\begin{equation*}
U(\phi)|\psi(\theta)\rangle
=|\psi(\theta-\phi)\rangle,
\quad
|\psi(\theta+2\pi)\rangle=|\psi(\theta)\rangle.
\end{equation*}
In particular, $e^{-iHt}|\psi(\theta)\rangle=|\psi(\theta-\omega t)\rangle$. Thus, the dynamical $U(1)$ structure is explicitly realized as an orbital phase-translation symmetry with coordinate $\theta$ parameterizing uniform motion along the effective orbit, irrespective of its shape in physical phase space.

These phase states are normalized vectors in the transformed Liouville space, rather than pure states of the original system. Their Fubini--Study line element is~\cite{ProvostVallee1980,Kolodrubetz2017}
\begin{equation}
ds_{\mathrm{FS}}^2=g_{\mathrm{FS}}\,d\theta^2,
\quad
g_{\mathrm{FS}}
=\langle\hat L^2\rangle-\langle\hat L\rangle^2
=\operatorname{Var}_{p}(l).
\label{eq:FSradius}
\end{equation}
Since the weights $p_l$ are independent of $\theta$, orbital phase translations act as isometries, and the Fubini--Study metric is homogeneous. The effective phase-state manifold therefore has the intrinsic geometry of a circle with radius $r_{\mathrm{FS}}=\sqrt{g_{\mathrm{FS}}}$, independently of the physical orbit's shape.

Figure~\ref{fig:limitCycle}(c) illustrates this homogeneous geometry, in contrast to the noncircular classical orbit and its quantum phase-space counterpart in Figs.~\ref{fig:limitCycle}(a) and (b). To examine the geometry of the density matrix, we reconstruct $|\rho(\theta)\rangle\rangle\propto\eta^{1/2}|\psi(\theta)\rangle$ with unit trace. For a valid density matrix $\rho(\theta)=\sum_a\lambda_a|a\rangle\langle a|$, the Bures metric is~\cite{BraunsteinCaves1994,PetzSudar1996}
\begin{equation}
g_{\mathrm B}(\theta)=\frac12\sum_{a,b} \frac{|\langle a|\partial_\theta\rho|b\rangle|^2}{\lambda_a+\lambda_b}.
\end{equation}
For an ideal quantum limit cycle, the contractivity of the Bures distance under completely positive trace-preserving evolution, together with periodicity, implies that $g_{\mathrm B}(\theta)$ is also homogeneous in the intrinsic orbital phase (see Appendix~\ref{sec:Reparameterization-App}). At finite size, the reconstructed Bures geometry exhibits weak angular modulation. 

The orbital structure is further characterized by the intrinsic phase distribution
\begin{equation}
P(\theta)=\int d^2\alpha Q(\alpha)
\delta(\theta-\theta(\alpha)),
\end{equation}
where $Q(\alpha)$ is the stationary Husimi function and $\theta(\alpha)$ is the orbital phase determined from $\theta(\alpha)=  \operatorname{Arg}[\langle \alpha| \tilde{\rho}^{\dagger}_1 |\alpha\rangle]-\theta_{0}$ with a reference phase $\theta_0$. As shown in Fig.~\ref{fig:limitCycle}(d), $P(\theta)$ is nearly uniform despite the noncircular phase-space structure in Fig.~\ref{fig:limitCycle}(b). Together, the effective Fubini--Study geometry, physical Bures geometry, and intrinsic phase statistics provide complementary characterizations of the underlying orbital $U(1)$ structure without requiring physical rotational symmetry. The correspondence between the effective quantum phase $\theta$ and the classical orbital phase is illustrated in Appendix~\ref{sec:Reparameterization-App}~\cite{Kato2019}.

\begin{figure}[t]
    \centering
    \includegraphics[width=\columnwidth]{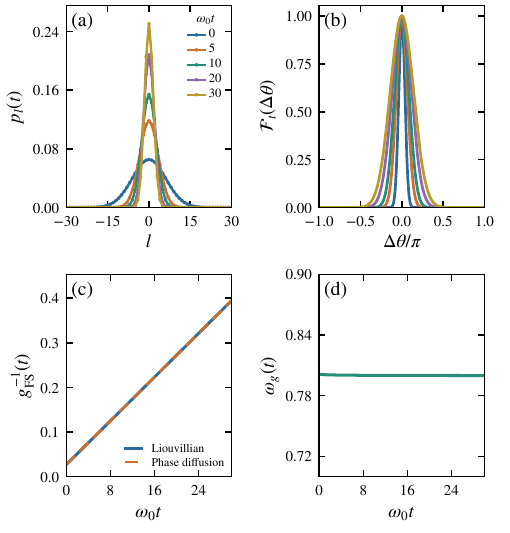}
    \caption{Effective phase-state geometry. (a) Normalized spectral weights $p_l(t)$ of the slow Liouvillian modes at selected times, showing the progressive suppression of higher angular-momentum harmonics. (b) Effective phase-state fidelity $\mathcal F_t(\Delta\theta)$, whose broadening reflects reduced phase distinguishability. (c) Inverse Fubini--Study metric $g_{\mathrm{FS}}^{-1}(t)$ calculated from the microscopic Liouvillian (solid line), compared with the quadratic phase-diffusion prediction (dashed line) using $\tilde D/\omega_0\simeq 0.0061$, independently extracted from the low-lying Liouvillian spectrum. (d) Geometric angular velocity $\omega_g(t)=\rm{Cov}_{p(t)}(l, \epsilon_l)/\rm{Var}_{p(t)}(l)$. Its nearly constant value is consistent with the approximately harmonic frequencies $\epsilon_l\simeq l\omega$.  Parameters are the same as in Fig.~\ref{fig:limitCycle}.}
    \label{fig:geometry}
\end{figure}

\section{Finite-size effect} For an ideal quantum limit cycle, a state $\rho(\theta_0)$ on the periodic orbit can be obtained from a coherent state $\rho_0=|\alpha_0\rangle\langle\alpha_0|$ localized near the classical limit cycle through stroboscopic long-time evolution, $\rho(\theta_0)=\lim_{n\to\infty}e^{nT\mathcal L}\rho_0$.  At finite size, however, the slow Liouvillian modes acquire small decay rates, causing the state to relax to the $\theta$-independent stationary state. To retain the orbital phase information, we instead evolve $\rho_0$ for a finite time $\tau$ and project it onto the selected slow Liouvillian sector,
\begin{equation}
|\rho_{\rm proj}\rangle\rangle
=\sum_l\mathcal P_l e^{\mathcal L\tau}|\rho_0\rangle\rangle.
\label{eq:refState}
\end{equation}
The preliminary evolution suppresses rapidly decaying transient components, while the spectral projection retains only the selected slow modes. Since the spectral projection does not necessarily preserve positivity, we reconstruct a positive-semidefinite, unit-trace reference state within the selected slow subspace, from which the phase-dependent states $|\rho(\theta)\rangle\rangle$ are constructed  (see Appendix~\ref{sec:Positivity-restore-App} for details). 

Keeping $\eta$ and the effective basis fixed,  we investigate how finite-size relaxation affects the effective $U(1)$ symmetry and Fubini--Study geometry. The slow Liouvillian eigenvalues take the form $z_l=-\gamma_l-i\epsilon_l$, with a small relaxation rate $\gamma_l\geq0$. The normalized effective phase states then evolve as
\begin{equation}
|\psi(\theta,t)\rangle =\sum_l\sqrt{p_l(t)} e^{il\theta-i\epsilon_lt}|l\rangle,
\end{equation}
where $p_l(t)=\frac{p_l(0)e^{-2\gamma_l t}} {\sum_m p_m(0)e^{-2\gamma_m t}}$ and $g_{\mathrm{FS}}=\rm{Var}_{p(t)}(l)$. As shown in Fig.~\ref{fig:geometry}(a), the fast decay of higher angular harmonics concentrates the spectral weight near $l=0$, contracting the effective geometric radius $r_{\mathrm{FS}}(t)=\sqrt{g_{\mathrm{FS}}(t)}$ shown in  Fig.~\ref{fig:geometry}(c).

The resulting loss of phase distinguishability is characterized by the fidelity
\begin{equation}
\mathcal F_t(\Delta\theta)=\left|\langle\psi(\theta,t)|\psi(\theta+\Delta\theta,t)\rangle\right|^2.
\label{eq:PhaseFidelity}
\end{equation}
Its curvature at $\Delta\theta=0$ determines the Fubini--Study metric $g_{\mathrm{FS}}(t)$. The broadening of the fidelity peak in Fig.~\ref{fig:geometry}(b) therefore reflects the progressive loss of phase distinguishability.

To characterize orbital phase evolution under finite-size relaxation, we introduce the geometric angular velocity
\begin{equation}
\omega_{\mathrm g}(t) \equiv-\frac{g_{\theta t}}{g_{\theta\theta}} =\frac{\operatorname{Cov}_{p(t)}(l,\epsilon_l)}{\operatorname{Var}_{p(t)}(l)},
\label{eq:GeometryVelocity}
\end{equation}
which measures the instantaneous orbital angular velocity.  For equally spaced frequencies $\epsilon_l=l\omega$, all harmonics rotate coherently, yielding a constant geometric angular velocity $\omega_{\mathrm g}(t)=\omega$. Otherwise, deviations from uniform frequency spacing can lead to a time-dependent angular velocity as the harmonic weights evolve.  As shown in Fig.~\ref{fig:geometry}(d), $\omega_{\mathrm g}(t)$ remains nearly constant, consistent with $\epsilon_l\simeq l\omega$. Thus, finite-size relaxation progressively reduces phase-state distinguishability while maintaining nearly uniform orbital phase evolution.  In the following section, we examine the physical origin of this symmetry-preserving relaxation.

\section{Quantum phase diffusion} The effective orbital $U(1)$ structure provides a natural framework for understanding finite-size relaxation, in close analogy with fluctuations in systems exhibiting spontaneous continuous-symmetry breaking~\cite{MinamiHidaka2018, Alaeian2021}. In the semiclassical regime, restoring dynamics suppresses the rapid transverse amplitude fluctuations, whereas the orbital phase remains neutrally stable. Under weak, short-correlated noise and a separation of amplitude and phase relaxation timescales, fluctuations along this neutral direction accumulate, leading to phase diffusion and a progressive loss of phase-state distinguishability, as illustrated in Fig.~\ref{fig:geometry}.

The coherent-state correspondence established in Appendix~\ref{sec:coherent-mapping-app} allows us to associate a complex amplitude with the effective phase states. Specifically, an annihilation-like operator $\mathscr A$ can be constructed such that $\mathscr A|\psi(\theta)\rangle=r_0e^{i\theta}|\psi(\theta)\rangle$, where $r_0=r_{\rm FS}(0)$ is the initial geometric radius (see Appendix~\ref{sec:effective-operators-app}).  This complex-amplitude representation provides a basis for modeling the slow amplitude and phase fluctuations using the stochastic Stuart--Landau equation~\cite{Nakao2016,KatoNakao2020,Lim2025},
\begin{equation}
\dot\psi=\bigl[-i\omega+\mu-g|\psi|^2\bigr]\psi+\eta(t),
\label{eq:StuartLandau}
\end{equation}
where $\psi=re^{-i\vartheta}$ and $\mu,g>0$. The deterministic dynamics supports a stable amplitude $r_0=\sqrt{\mu/g}$ and a freely rotating phase, $\dot\vartheta=\omega$.

Assuming weak Gaussian white noise, the leading-order radial and angular dynamics are
\begin{align}
    \dot r=\mu r-gr^3+\eta_r(t),~~ \dot\vartheta=\omega+\frac{\eta_\vartheta(t)}{r}.
    \label{eq:RadialPhase}
\end{align}
where $\langle\eta_i(t)\eta_j(t')\rangle=D_i\delta_{ij}\delta(t-t')$ for $i=r,\vartheta$. Linearizing around the stable amplitude $r_{0}=\sqrt{\mu/g}$ gives a radial relaxation rate $\gamma_r=2\mu$. When radial relaxation is much faster than phase diffusion, we may approximate $r\simeq r_{0}$, yielding an effective phase-diffusion coefficient $\tilde D\simeq D_\vartheta/r_{0}^2$. Thus, for fixed noise strength and radial restoring rate, this separation becomes increasingly accurate in the large-amplitude regime when $r_{0}\gg 1$.

On these coarse-grained timescales, the phase dynamics can be represented by the stochastic generator
\begin{equation}
    H(t)=-i[\omega+\xi(t)]
    \partial_\theta,
    \quad
    \langle\xi(t)
    \xi(t')\rangle
    =\tilde D \delta(t-t'),
    \label{eq:StochasticHamiltonian}
\end{equation}
where $\xi(t)$ denotes effective Gaussian white noise. Averaging over its realizations yields quadratic damping of the angular harmonics,
\begin{equation}
\mathbb{E}_{\xi}\left[e^{-i\int_0^t H(\tau) d\tau}
\right]|l\rangle=e^{-il\omega t-\gamma_l t}|l\rangle,
\quad \gamma_l=\frac{\tilde D}{2}l^2.
    \label{eq:DiffusiveSpectrum}
\end{equation}
Thus, phase diffusion damps the oscillatory Liouvillian modes while preserving the effective $U(1)$ phase-translation covariance and nearly uniform orbital phase rotation.

\begin{figure}[t]
    \centering
    \includegraphics[width=\columnwidth]{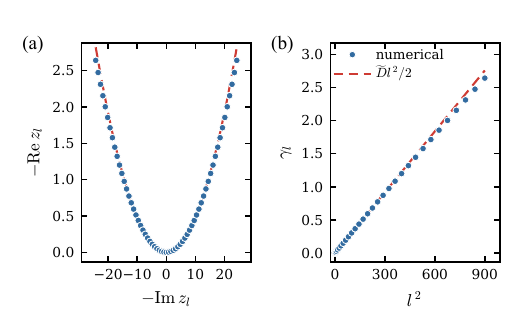}
    \caption{ Slow Liouvillian spectrum and phase diffusion. (a) Complex slow-mode eigenvalues
    $z_l=-\gamma_l-i\epsilon_l$, with $\epsilon_l\simeq l\omega$. (b) Decay rates $\gamma_l$ versus $l^2$, compared with the quadratic prediction $\gamma_l=\tilde D l^2/2$ (dashed line). Fitting $1\leq l\leq5$ gives $\tilde D/\omega_0\simeq0.0061$. Parameters are the same as in Fig.~\ref{fig:limitCycle}.
    }
    \label{fig:DiffusionEffect}
\end{figure}

Figure~\ref{fig:DiffusionEffect} compares this prediction with the slow Liouvillian spectrum. The approximately parabolic dispersion of the decay rates is consistent with phase diffusion. Fitting the low-lying harmonics yields $\tilde D/\omega_0\simeq0.0061$, while deviations from quadratic scaling become more pronounced at larger $|l|$, where the separation between radial relaxation and phase diffusion breaks down.

The same diffusive spectrum predicts the contraction of the effective phase-state geometry. Since each angular harmonic decays as $e^{-\tilde D l^2t/2}$, its normalized spectral weight evolves as
\begin{equation}
p_l(t)= \frac{p_l(0)e^{-\tilde D l^2t}} {\sum_m p_m(0)e^{-\tilde D m^2t}}.
\label{e:DiffusiveWeights}
\end{equation}
Substituting these weights into $g_{\rm FS}(t)=\operatorname{Var}_{p(t)}(l)$ yields the geometric contraction predicted by phase diffusion, as shown in Fig.~\ref{fig:geometry}(c). The agreement with the microscopic Liouvillian results establishes a quantitative connection between finite-size phase diffusion and the loss of distinguishability of effective orbital phase states.

\section{Discussion and conclusion} We have recovered a unitary representation of the underlying orbital $U(1)$ symmetry of quantum limit cycles without requiring rotational symmetry in physical phase space. By mapping the harmonic Liouvillian sector onto a Hermitian angular-momentum Hamiltonian, we identify dynamical time evolution with orbital phase translations. The resulting phase states possess homogeneous Fubini--Study geometry. This orbital symmetry also provides a natural framework for understanding finite-size phase diffusion, in close analogy with fluctuations in systems exhibiting spontaneous continuous-symmetry breaking. In the semiclassical weak-noise regime, fluctuations accumulate along the neutrally stable phase direction while transverse deviations rapidly relax, leading to diffusive damping of the slow Liouvillian harmonics. Such relaxation preserves the effective orbital $U(1)$ covariance while progressively reducing phase-state distinguishability, with nearly uniform phase evolution. The microscopic slow-mode spectrum quantitatively predicts the resulting contraction of the effective geometry. Our results thus extend the spectral characterization of quantum limit cycles into a symmetry-based geometric framework connecting orbital dynamics, quantum geometry, and finite-size phase diffusion.

Our construction also suggests extensions beyond periodic limit cycles. For a diagonalizable nondecaying Liouvillian sector with frequencies that are not necessarily equally spaced, the same similarity transformation yields an effective Hermitian Hamiltonian,
\begin{equation}
H=\sum_j\epsilon_j|j\rangle\langle j|.
\end{equation}
When the spectrum contains multiple incommensurate fundamental frequencies, the dynamics can become quasiperiodic, suggesting a generalization from the orbital phase circle to a higher-dimensional phase manifold~\cite{KatoNakao2023}. Its geometry may be characterized by the Fubini--Study metric of the corresponding effective states.

A more challenging extension concerns dissipative systems with classical strange attractors, as encountered in nonlinear laser dynamics~\cite{Haken1975}.  Its complex signatures may require the spectral structure and statistics of slowly decaying modes and their semiclassical correspondence~\cite{Dutta2025}. Whether the present geometric framework can reveal connections between Liouvillian spectral statistics and classical chaos, analogous to those underlying the Bohigas--Giannoni--Schmit conjecture~\cite{Bohigas1984}, remains an open question.

Our results thus establish a unified connection between emergent orbital symmetry, quantum geometry, and finite-size phase diffusion, providing a foundation for exploring more general nonequilibrium dynamics in open quantum systems.

\section{Acknowledgments}
 We thank Dr. Da-Wu Xiao for helpful discussions.  This work was supported by the Innovation Program for Quantum Science and Technology Project No. 2023ZD0300600, the Hong Kong Research Grants Council - General Research Fund (Project No. 14302121), the National Natural Science Foundation of China/Hong Kong Research Council Collaborative Research Scheme Project CRS-CUHK401/22, the Hong Kong Research Grants Council Senior Research Fellow Scheme Project SRFS2223-4S01, and the New Cornerstone Science Foundation.

\appendix

\renewcommand{\section}[1]{%
    \refstepcounter{section}%
    \begin{center}
        \normalfont\bfseries
        Appendix \Alph{section}:
        #1
    \end{center}
}

\section{Gaussian regularization of the reference state}\label{sec:Positivity-restore-App}

For an ideal quantum limit cycle, projection onto the complete nondecaying Liouvillian sector preserves positivity. Indeed, for a periodic orbit with period $T=2\pi/\omega$, the projection can be expressed as
\begin{equation}
\mathcal P_{\mathrm{DFS}}\rho_0
=\lim_{n\to\infty}e^{nT\mathcal L}\rho_0.
\label{eq:IdealProjection}
\end{equation}
Since $e^{t\mathcal L}$ is completely positive and trace preserving, the projected state remains a density matrix.

\begin{figure}
    \centering
    \includegraphics[width=1.0\linewidth]{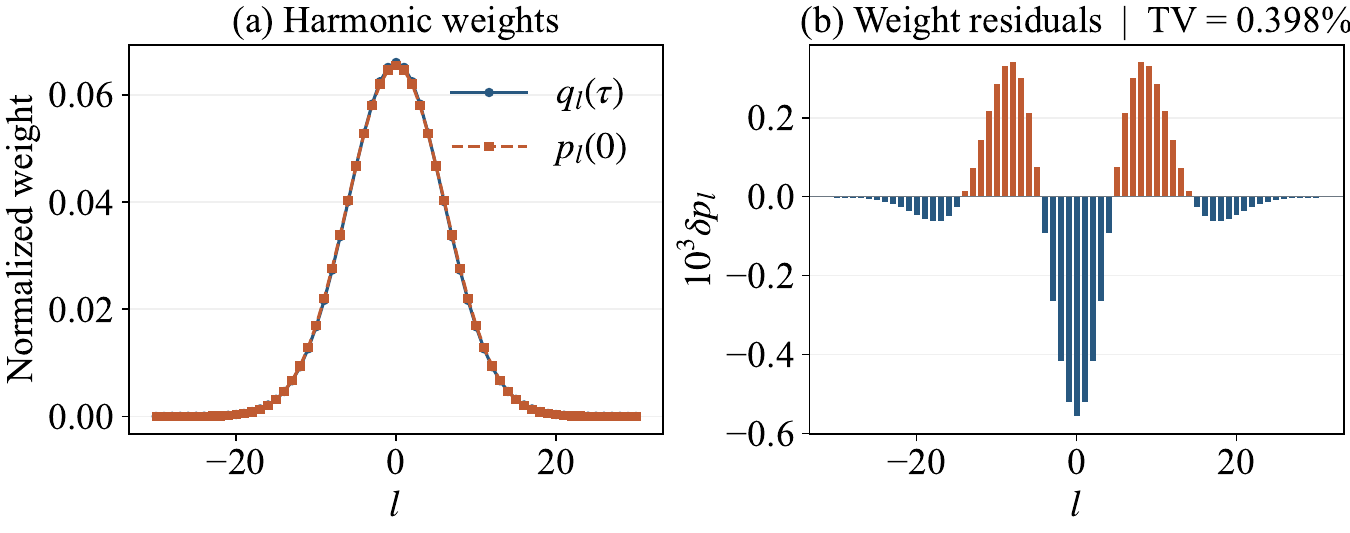}
    \caption{Comparison of the projected harmonic weights $q_l(\tau)$
and their Gaussian approximation $p_l(0)$ for $\omega_0\tau=1.5$.
(a) Normalized harmonic weights for the evolved projection and the
Gaussian reference. (b) Residuals $\delta p_l=p_l(0)-q_l(\tau)$.
The total variation distance $\frac{1}{2}\sum_l|\delta p_l|$ is
$0.398\%$. The model parameters are the same as in the main text.}
    \label{fig:GaussianFixing-app}
\end{figure}

For finite-size system, we choose a coherent state $\rho_0=|\alpha_0\rangle\langle\alpha_0|$
at the point of maximum radius on the settled mean-field limit cycle, with $\alpha_0=(x_0+i p_0)/\sqrt{2}$. We evolve $\rho_0$ for a time $\tau$ and project it onto the selected slow Liouvillian modes,
\begin{equation}
    |\rho_{\mathrm{proj}}(\tau)\rangle\rangle
    =\sum_l \mathcal{P}_l e^{\mathcal{L}\tau}|\rho_0\rangle\rangle
    =\sum_l c_l e^{z_l\tau}|R_l\rangle\rangle ,
\end{equation}
where $z_l$ is the eigenvalue of the Hilbert--Schmidt-normalized right
mode $R_l$. The normalized harmonic weights of the evolved projection
are
\begin{equation}
    q_l(\tau)=
    \frac{|c_l|^2 e^{2\tau\operatorname{Re}z_l}}
         {\sum_m |c_m|^2 e^{2\tau\operatorname{Re}z_m}} .
\end{equation}

The distribution is approximated by a centered Gaussian on the same set of harmonics,
\begin{equation}
    p_l(0)=
    \frac{\exp[-l^2/(2\sigma^2)]}
         {\sum_m \exp[-m^2/(2\sigma^2)]} ,
\end{equation}
by choosing the trial width $\sigma$ to satisfy
\begin{equation}
    \operatorname{Var}_{p(0)}(l)
    =\operatorname{Var}_{q(\tau)}(l).
\end{equation}
Finite-size phase diffusion motivates the Gaussian ansatz because
quadratic decay rates in $l$ generate a Gaussian suppression of the high harmonics.
The accuracy of the approximation depends on the resulting
distribution. If the target variance exceeds the uniform-distribution
variance on the retained harmonics, we use the uniform limit
$\sigma\to\infty$ as the trial profile.

We construct the Gaussian reference from the projected state at time
$\tau$. By that time, each mode coefficient has evolved from $c_l$ to
$c_l e^{z_l\tau}$. We replace its magnitude with the Gaussian amplitude
$|c_0|e^{-l^2/(4\sigma^2)}$ and retain its phase
$\arg(c_l)+\operatorname{Im}(z_l)\tau$ whenever $c_l\ne 0$.
The resulting state defines $\theta=0$. An orbital shift by $\theta$
then multiplies the coefficient of harmonic $l$ by $e^{il\theta}$. If $\operatorname{Im}(z_l)=l\omega$, the phase accumulated during
preparation is $l\omega\tau$ for each harmonic. In this case, the
preparation shifts the orbital phase origin by $\omega \tau$.

We accept the trial width if the reconstructed Gaussian satisfies
$\lambda_{\min}\geq-\varepsilon_{\mathrm{PSD}}$ at the reference phase and the selected observation times. If the test fails, we reduce the width until it passes and report the remaining variance mismatch. In calculation, $\varepsilon_{\mathrm{PSD}}=10^{-4}$ is used. 

Figure~\ref{fig:GaussianFixing-app} compares the projected distribution
$q_l(\tau)$ with the accepted Gaussian profile $p_l(0)$ for
$\omega_0\tau=1.5$. At the reference phase, the smallest eigenvalue
increases from $-1.1\times10^{-4}$ for the evolved projection to
$-6.17\times10^{-5}$ for the Gaussian reconstruction.

\section{Reparameterization of the limit cycle}\label{sec:Reparameterization-App}

Starting from the mean-field equations, we numerically determine the
classical limit-cycle trajectory $(x(t),p(t))$ in phase space,
as shown in Fig.~\ref{fig:limitCycle}(a). Introducing the polar
coordinates $x=r\cos\phi$ and $p=r\sin\phi$, the mean-field
equations become
\begin{subequations}
\begin{align}
    \dot r &=
    \left(
    \frac{\gamma_\uparrow-\gamma_\downarrow}{2}
    -\frac{\kappa}{2}r^2
    -\epsilon\sin 2\phi
    \right)r,\\
    \dot\phi &=
    -(\omega_0+\epsilon\cos 2\phi).
\end{align}
\end{subequations}
For $\omega_0>|\epsilon|$, the angular motion is monotonic,
and the oscillation period can be obtained analytically as
\begin{equation}
    T=
    \int_0^{2\pi}
    \frac{d\phi}{\omega_0+\epsilon\cos 2\phi}
    =
    \frac{2\pi}{\sqrt{\omega_0^2-\epsilon^2}}.
\end{equation}
Although the physical angular velocity depends on $\phi$, a uniformly rotating effective phase can be introduced through~\cite{Kato2019}
\begin{equation}
    \label{eq:theta-app}
    \theta_{\rm lc}(\phi)
    =
    \omega
    \int_0^\phi
    \frac{d\varphi}{\omega_0+\epsilon\cos 2\varphi},
\end{equation}
where $\omega=\sqrt{\omega_0^2-\epsilon^2}$.
By construction, this phase satisfies
\begin{equation}
    \label{eq:thetaMotion-app}
    \dot{\theta}_{\rm lc}=-\omega.
\end{equation}
Thus, despite the noncircular shape of the limit cycle,
its dynamics admits a homogeneous phase coordinate,
providing a natural classical realization of $U(1)$
phase translations.

\begin{figure}
    \centering
    \includegraphics[width=1.0\linewidth]{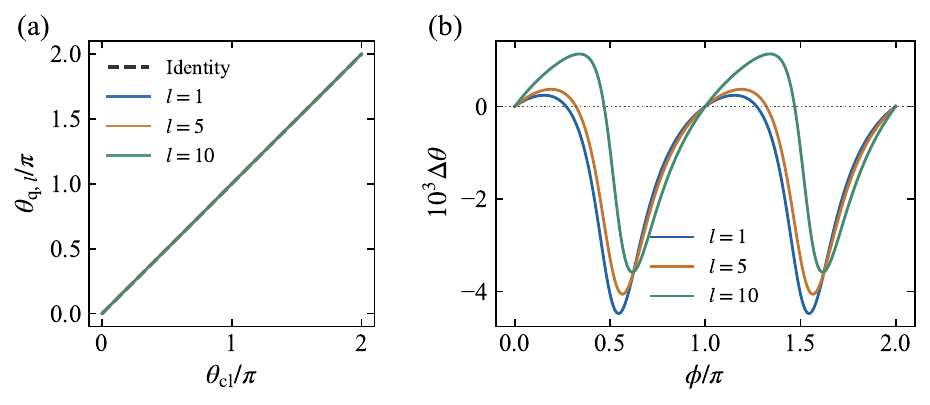}
    \caption{Correspondence between the classical and quantum effective phases (a) Quantum phase $\theta_{\mathrm q,l}\equiv \arg[c_l(\phi)/c_l(0)]/l$, extracted from the projections of coherent states along the classical limit cycle, plotted against the classical effective phase $\theta_{\mathrm{lc}}$. The solid black line indicates $\theta_{\mathrm q,l}=\theta_{\mathrm{lc}}$. (b) Phase deviation $\Delta\theta=\theta_{\mathrm q,l}-\theta_{\mathrm{lc}}$ as a function of the polar angle $\phi$, for $l=1,5,10$. The close agreement demonstrates the correspondence between the quantum phase parameter and the uniformly rotating classical phase, with small harmonic-dependent deviations.}
    \label{fig:theta-app}
\end{figure}

We next establish the correspondence between $\theta_{\rm lc}$ and the phase parameter $\theta$ in the effective quantum state $|\psi(\theta)\rangle$.
Consider a coherent state $\rho_\phi=|\alpha_\phi\rangle\langle\alpha_\phi|$
localized on the classical limit cycle, with
\begin{equation}
    \alpha_\phi=
    \frac{x_c(\phi)+ip_c(\phi)}{\sqrt{2}},
    \qquad \phi=\arg\alpha_\phi.
\end{equation}
The classical effective phase associated with this
state is $\theta_{\rm lc}(\phi)$, as defined in
Eq.~\eqref{eq:theta-app}. Independently, the quantum phase can be extracted from the projections of $\rho_\phi$ onto the left
slow Liouvillian eigenmodes~\cite{KatoNakao2022},
\begin{equation}
    c_l(\phi) =\langle \langle\tilde{\rho}_l |\rho_\phi\rangle\rangle.
\end{equation}
After fixing the eigenmode gauge such that
$c_l(0)$ is real and positive, the phase associated
with the $l$th harmonic is
\begin{equation}
    \theta_l(\phi)
    =\frac{1}{l}\arg c_l(\phi).
\end{equation}
In the semiclassical limit, when the left slow
Liouvillian eigenmodes approach the classical
phase harmonics, their projections satisfy
\begin{equation}
    c_l(\phi)\simeq
    A_l e^{il\theta_{\rm lc}(\phi)},
\end{equation}
where $A_l>0$ is independent of $\phi$ to leading order. Consequently,
\begin{equation}
    \theta_l(\phi)
    \simeq\theta_{\rm lc}(\phi)
\end{equation}
for the resolved harmonics, with a consistent continuous phase unwrapping.  Numerical results of $\theta_{\rm lc}$ and $\theta_{\rm{q},l}$ and their difference are shown in Fig.~\ref{fig:theta-app} (a) and (b), respectively, which verify the corresponding relation between $\theta_{\rm lc}$ and $\theta_{\rm{q},l}$. 

The homogeneous character of the effective phase can also be examined through the stationary state $\rho_s$, particularly its Husimi distribution
\begin{equation}
    Q(\alpha) =\frac{1}{\pi} \langle\alpha|\rho_s|\alpha\rangle,
\end{equation}
as shown in Fig.~\ref{fig:limitCycle}(b) of the main text. Here, $|\alpha\rangle$ denotes a coherent state with $\alpha=(x+ip)/\sqrt{2}$. The Husimi function is nonnegative and normalized according to $\int d^2\alpha\,Q(\alpha)=1$.

Writing $\alpha=re^{i\phi}/\sqrt{2}$, we obtain the marginal distribution of the polar angle,
\begin{equation}
    P(\phi)= \frac{1}{2}
    \int_0^\infty r\,dr  Q\left(\frac{r} {\sqrt{2}}e^{i\phi}\right).
    \label{eq:Pphi-app}
\end{equation}
The corresponding distribution of the effective phase $\theta$ is obtained by the transformation $\theta=\theta_{l}(\phi)$,
\begin{equation}
    P(\theta)= \int_0^{2\pi}d\phi\, P(\phi) \delta (\theta-\theta_{l}(\phi)).
    \label{eq:Ptheta-app}
\end{equation}

\begin{figure}[t]
    \centering
    \includegraphics[width=0.8\linewidth]{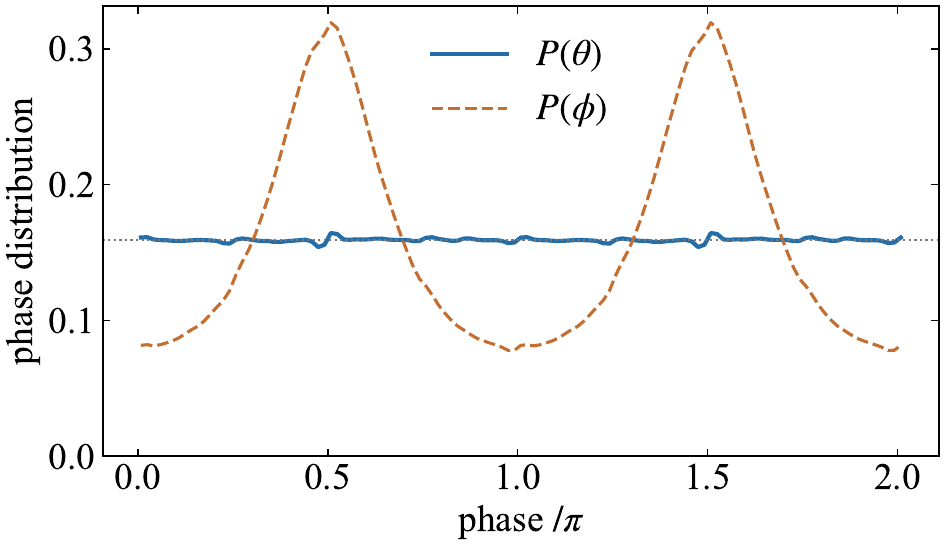}
    \caption{
    Stationary phase distributions obtained from
    the Husimi function $Q(\alpha)$.
    The polar-angle distribution $P(\phi)$ is calculated using Eq.~\eqref{eq:Pphi-app}, while the effective-phase distribution $P(\theta)$ follows from Eq.~\eqref{eq:Ptheta-app}.
    The nearly uniform $P(\theta)$ contrasts with the strongly anisotropic $P(\phi)$, illustrating the homogeneous stationary
    measure in the effective phase coordinate.
    }
    \label{fig:Q-distribution-app}
\end{figure}

Figure~\ref{fig:Q-distribution-app} compares $P(\phi)$ and $P(\theta)$ extracted from the stationary quantum Husimi distribution. While $P(\phi)$ is strongly nonuniform,
$P(\theta)$ is nearly constant and closely approaches $1/(2\pi)$.
This agreement demonstrates that the effective phase transformation removes the dominant kinematic anisotropy associated with the nonuniform physical angular velocity. The resulting homogeneous phase statistics are consistent with the effective $U(1)$ phase-translation structure identified
through the slow Liouvillian modes. 

For an ideal quantum limit cycle, the Bures geometry of the density matrix is also homogeneous. Consider a periodic family $\rho_\theta\equiv\rho(\theta)$ satisfying $\Phi_t[\rho_\theta]=\rho_{\theta-\omega t}$, where $\Phi_t=e^{t\mathcal L}$ is a completely positive trace-preserving map and $T=2\pi/\omega$ is the oscillation period. The contractivity of the Bures distance implies
\begin{equation}
D_{\mathrm B}(\rho_{\theta-\omega t},\rho_{\theta+\delta-\omega t})
\leq D_{\mathrm B}(\rho_\theta,\rho_{\theta+\delta}).
\end{equation}
Evolving both states for the remaining period $T-t$ yields the reverse inequality by periodicity. Thus, the Bures distance is invariant under orbital phase translations, implying
\begin{equation}
g_{\mathrm B}(\theta)=\mathrm{const}.
\label{eq:BuresHomogeneity}
\end{equation}
Hence, an ideal quantum limit cycle possesses homogeneous Bures geometry regardless of its shape in physical phase space. At finite size, the reconstructed Bures metric exhibits weak angular modulation [Fig.~\ref{fig:limitCycle}(c)], reflecting deviations from ideal periodic dynamics, although reconstruction and coordinate effects may also contribute.

\section{Correspondence between effective phase states and coherent states}
\label{sec:coherent-mapping-app}
To clarify the physical meaning of the Hermitian representation, we compare it with a family of coherent-state density matrix whose phase varies at fixed amplitude,
\begin{equation}
    |\alpha(\theta)\rangle\rangle
    \equiv|\alpha(\theta)\rangle\langle\alpha(\theta)|,
    \qquad \alpha(\theta)=Re^{i\theta}.
    \label{eq:liouville-coherent}
\end{equation}
Here, $|\alpha(\theta)\rangle\rangle$ denotes a density matrix represented as a vector in Liouville space. Expanding it in the Fock basis gives
\begin{equation}
    |\alpha(\theta)\rangle\rangle
    =e^{-R^2}\sum_{n,m=0}^{\infty}
    \frac{R^{n+m}e^{i(n-m)\theta}}{\sqrt{n!m!}}
    |n,m\rangle\rangle,
\end{equation}
where $|n,m\rangle\rangle=\bigl||n\rangle\langle m|\bigr\rangle\rangle$. Unlike a coherent state in the original Hilbert space, its Liouville-space representation contains both positive and negative angular harmonics, labeled by $l=n-m\in\mathbb Z$.

To separate these harmonics, we introduce the Fourier components
\begin{equation}
    |\rho_l^C\rangle\rangle
    =\frac{1}{2\pi}\int_0^{2\pi}d\theta\,
    e^{-il\theta}|\alpha(\theta)\rangle\rangle.
    \label{eq:coherent-fourier}
\end{equation}
Different harmonics are orthogonal under the Hilbert--Schmidt inner product, with squared norms
\begin{equation}
    q_l=\langle\langle\rho_l^C|\rho_l^C\rangle\rangle
    =e^{-2R^2}I_{|l|}(2R^2),
    \label{eq:coherent-harmonic-weights}
\end{equation}
where $I_l$ is the modified Bessel function of the first kind. Defining the normalized harmonic modes $|l\rangle_C=|\rho_l^C\rangle\rangle/\sqrt{q_l}$, we obtain
\begin{equation}
    |\alpha(\theta)\rangle\rangle
    =\sum_{l=-\infty}^{\infty}
    \sqrt{q_l}\,e^{il\theta}|l\rangle_C.
    \label{eq:coherent-liouville-expansion}
\end{equation}
This expansion has the same phase dependence as $|\psi(\theta)\rangle$ defined in the main text, as both families are superpositions of integer angular harmonics with phase factors $e^{il\theta}$, although their harmonic weights and basis states generally differ.

The correspondence also provides a geometric interpretation of the effective radius. For the coherent-state density matrix in Eq.~\eqref{eq:liouville-coherent}, the Fubini--Study metric in Liouville space is
\begin{equation}
    g_{\rm FS}^{C}
    =\operatorname{Var}_{q}(l)
    =2R^2.
    \label{eq:liouville-coherent-metric}
\end{equation}
The factor of two arises from representing pure-state density matrix rather than state vectors. Since the Bures metric of pure states coincides with the Fubini--Study metric in the original Hilbert space, we have
\begin{equation}
    g_{\rm FS}^{C}=2g_{\rm B}=2R^2.
\end{equation}
Choosing $R^2=r_{\rm FS}^2/2$ gives
\begin{equation}
    g_{\rm FS}^{C}=g_{\rm FS}=r_{\rm FS}^2.
\end{equation}
Thus, the two phase-state manifolds are intrinsically isometric, although their harmonic weights and finite-angle overlaps generally differ. This correspondence identifies the effective geometric radius with the amplitude scale of an equivalent coherent-state representation, providing a geometric interpretation of the complex amplitude introduced in the main text.

\section{Effective annihilation and number operators}
\label{sec:effective-operators-app}

The coherent-state correspondence motivates the introduction of an effective annihilation-like
operator whose eigenvalue defines a complex coordinate on the phase-state manifold.
Consider the effective phase states
\begin{equation}
    |\psi(\theta)\rangle
    =\sum_l\sqrt{p_l}\,e^{il\theta}|l\rangle,
    \qquad
    \hat L|l\rangle=l|l\rangle,
\end{equation}
where $\sum_l p_l=1$. Their Fubini--Study metric is determined by the angular-momentum
fluctuations,
\begin{equation}
    g_{\rm FS}
    =\langle\hat L^2\rangle
    -\langle\hat L\rangle^2
    \equiv r_{\rm FS}^2.
\end{equation}

We introduce an effective annihilation operator
\begin{equation}
    \mathscr A
    =r_{\rm FS}\sum_l
    \sqrt{\frac{p_{l-1}}{p_l}}\,
    |l-1\rangle\langle l|,
    \label{eq:effective-annihilation}
\end{equation}
assuming an untruncated harmonic representation with nonvanishing $p_l$. By construction, it satisfies
\begin{equation}
    \mathscr A|\psi(\theta)\rangle
    =r_{\rm FS}e^{i\theta}|\psi(\theta)\rangle,
    \qquad
    [\hat L,\mathscr A]=-\mathscr A.
    \label{eq:effective-annihilation-eigen}
\end{equation}
The eigenvalue $r_{\rm FS}e^{i\theta}$ therefore provides a complex coordinate on the effective
phase-state manifold, analogous to the coherent-state amplitude of a harmonic oscillator.

The corresponding effective number operator is defined by
\begin{equation}
    \mathscr N
    =\mathscr A^\dagger\mathscr A
    =r_{\rm FS}^2\sum_l
    \frac{p_{l-1}}{p_l}|l\rangle\langle l|.
    \label{eq:effective-number}
\end{equation}
It is positive and commutes with the angular-momentum operator,
\begin{equation}
    [\mathscr N,\hat L]=0.
\end{equation}
Furthermore, Eq.~\eqref{eq:effective-annihilation-eigen} implies
\begin{equation}
    \langle\psi(\theta)|\mathscr N
    |\psi(\theta)\rangle=r_{\rm FS}^2.
    \label{eq:effective-number-expectation}
\end{equation}
Thus, $\mathscr N$ provides an occupation-like interpretation of the squared effective radius.

To establish its relation to the angular-momentum operator, we consider the semiclassical regime
$r_{\rm FS}^2\gg1$, in which the harmonic distribution is assumed to be approximately Gaussian, given by
\begin{equation}
    p_l\simeq \frac{1}{\sqrt{2\pi r_{\rm FS}^2}} \exp\left(-\frac{l^2}{2r_{\rm FS}^2}\right).
    \label{eq:gaussian-harmonics}
\end{equation}
Here, the distribution is centered at $l=0$, with variance $r_{\rm FS}^2$. It follows that
\begin{equation}
    \frac{p_{l-1}}{p_l}
    \simeq
    \exp\left(\frac{l-\frac12}{r_{\rm FS}^2}\right).
\end{equation}
Consequently, the effective number operator takes the approximate form
\begin{equation}
    \mathscr N \simeq r_{\rm FS}^2\exp\left(\frac{\hat L-\frac12}{r_{\rm FS}^2} \right).
    \label{eq:effective-number-exponential}
\end{equation}

Since the dominant harmonics satisfy $|l|\sim r_{\rm FS}$, the exponent is of order
$1/r_{\rm FS}$ and can be expanded in the semiclassical limit:
\begin{equation}
    \mathscr N \simeq r_{\rm FS}^2+\hat L+\mathcal O\left(1\right).
\label{eq:effective-number-expansion}
\end{equation}

This result provides a semiclassical interpretation of the shifted angular-momentum operator as an effective action variable. In particular, the complex amplitude $re^{i\theta}$ may be associated with the effective annihilation operator $\mathscr A$, while its squared magnitude
is represented by $\mathscr N$. This construction motivates an effective stochastic Stuart--Landau description, in which radial restoring dynamics and phase fluctuations govern the evolution of the complex amplitude. However, the radial dynamics and noise coefficients
are not determined by the phase-state geometry alone and must be obtained from
the underlying Liouvillian dynamics.

We emphasize that the relation $\mathscr N\simeq I_0+\hat L$ is valid only within the central harmonic region in the semiclassical regime. The exact operator $\mathscr N=\mathscr A^\dagger\mathscr A$ remains positive, whereas the shifted angular-momentum operator need not be
positive over the entire harmonic spectrum. Moreover, $\mathscr A$ is generally a
weighted-shift operator and need not satisfy the canonical bosonic commutation
relation.

\bibliography{quantumLimitCycle}

\end{document}